*Ab initio investigation of layered TMGeTe$_3$ alloys for phase-change applications*

Yihui Jiang[1,2], Suyang Sun[2,3,*], Hanyi Zhang[2], Xiaozhe Wang[2], Yibo Lei[1], Riccardo Mazzarello[4,*], Wei Zhang[2,*]

[1]Key Laboratory of Synthetic and Natural Functional Molecule of the Ministry of Education, College of Chemistry & Materials Science, Shaanxi key Laboratory of Physico-Inorganic Chemistry, Northwest University, Xi'an 710127, China.
[2]Center for Alloy Innovation and Design (CAID), State Key Laboratory for Mechanical Behavior of Materials, Xi'an Jiaotong University, Xi'an 710049, China.
[3]Institute of Materials, Henan Academy of Sciences, Zhengzhou 450046, China
[4]Department of Physics, Sapienza University of Rome, Rome 00185, Italy.

*Emails: sy.sun@hnas.ac.cn, riccardo.mazzarello@uniroma1.it, wzhang0@mail.xjtu.edu.cn

**Abstract:**
Chalcogenide phase-change materials (PCMs) are one of the most mature candidates for next-generation memory technology. Recently, CrGeTe$_3$ (CrGT) emerged as a promising PCM due to its enhanced amorphous stability and fast crystallization for embedded memory applications. The amorphous stability of CrGT was attributed to the complex layered structure of the crystalline motifs needed to initiate crystallization. A subsequent computational screening work identified several similar compounds with good thermal stability, such as InGeTe$_3$, CrSiTe$_3$ and BiSiTe$_3$. Here, we explore substitution of Cr in CrGT with other 3$d$ metals, and predict four additional layered alloys to be dynamically stable, namely, ScGeTe$_3$, TiGeTe$_3$, ZnGeTe$_3$ and MnGeTe$_3$. Thorough *ab initio* simulations performed on both crystalline and amorphous models of these materials indicate the former three alloys to be potential PCMs with sizable resistance contrast. Furthermore, we find that crystalline MnGeTe$_3$ exhibits ferromagnetic behavior, whereas the amorphous state probably forms a spin-glass phase. This makes MnGeTe$_3$ a promising candidate for magnetic phase-change applications.

# 1. Introduction

Chalcogenide phase-change materials (PCM), which combine fast read and write operations with non-volatile storage, are one of the most mature candidates for the next-generation memory technology[1-12]. The large electrical resistance contrast between the crystalline phase (logic "1") and the amorphous phase (logic "0") is utilized to store binary digits in memory units. The flagship PCMs along the GeTe–$Sb_2Te_3$ pseudo-binary line[13-15] show a balanced materials portfolio, including rapid phase transition capacity, fairly good thermal stability, sizable property contrast as well as high cyclability. In particular, slightly doped $Ge_2Sb_2Te_5$ (GST) has enabled massive production of high-density persistent memory cards[16]. The sizable resistance contrast between amorphous and crystalline GST can also accommodate many intermediate resistance levels for in-memory computing technology[17-24]. However, the relatively low crystallization temperature ($T_x$) of pure GST around 150 °C is a hindrance to embedded memory applications, and heavy doping (alloying) is needed to increase $T_x$. Indeed, heavy doping GST with carbon[25-27] or germanium (the latter is also called Ge-rich GST)[28-32] can effectively enhance the $T_x$ to 260 °C and above. Nevertheless, this off-stoichiometric alloying approach also increases the probability of phase separation. Thus, new PCMs with intrinsically high amorphous stability are being pursued.

A van der Waals layered compound $CrGeTe_3$ (CrGT) was recently investigated for phase-change memory applications[33-39], which shows a very good amorphous stability with $T_x$ ~276 °C. However, the resistance contrast window between amorphous and crystalline CrGT spans only one order of magnitude, much smaller than that of GST spanning three orders of magnitude. Very recently, we have carried out a computational materials screening[40] over the inorganic crystal structure database, and identified six other tellurides structurally similar to CrGT, namely, $InGeTe_3$ (InGT), $CrSiTe_3$, $ScSiTe_3$, $BiSiTe_3$, $InSiTe_3$ and $AlSiTe_3$. We revealed that these alloys exhibit consistently higher $T_x$ than traditional PCMs due to the complexity of nucleation into two-dimensional seeds out of a three-dimensional amorphous phase. In particular, InGT shows an intrinsic $T_x$ of ~260 °C and also a much larger resistance contrast window over five orders of magnitude[40]. We note that, except for CrGT and InGT, the other identified alloys contain silicon atoms rather than germanium atoms. High concentration of silicon can slow down the crystallization speed at elevated temperatures and increases the possibility of phase separation.

In parallel to materials screening over existing materials databases[40-43], chemical substitution has also proven to be a useful approach for developing new PCMs[44-46]. In this work, we focus on the substitution of chromium atoms in CrGT with other 3$d$ transition metal elements, and determine whether some of them can be used for phase-change memory applications. The presence of high amount of germanium and tellurium atoms should guarantee relatively fast crystallization at elevated temperatures. Another advantage of performing chemical substitution on the transition metal site is that we can modify their magnetic properties, which opens up another degree of freedom for phase-change tuning[47,48]. In an earlier work, we revealed the importance of spin-polarization in stabilizing the resistance contrast window in CrGT[39]. In addition, we proved that amorphous (a-) CrGT forms a spin glass state below ~20K. Taking into account the ferromagnetic behavior of crystalline (c-) CrGT up to ~66 K[49], it is in principle possible to achieve fast magnetic switching[47] at very low temperatures by programming CrGT with external electrical or laser pulses. It remains elusive whether the substitution with other 3$d$ transition metal elements could also lead to compounds with interesting magnetic features.

## 2. Results and discussion

Figure 1 illustrates the atomic structure of CrGT[50], and by replacing the center Cr atoms with other 3$d$ transiton metal (TM) atoms, we created a series of TMGeTe$_3$ (TMGT) crystals. The TM, Ge and Te atoms are rendered in purple, orange and green spheres respectively. In each atomic slab, TM atoms are coordinated with six Te forming edge-sharing TM[Te$_6$] octahedra, and Ge atoms are bonded to three Te atoms and one Ge atom, forming Ge[GeTe$_3$] tetrahedral motifs. These motifs are represented as red and blue polyhedra respectively. The atomic positions and lattice parameters of all ten TMGT crystals were optimized using density functional theory (DFT) with the Perdew-Burke-Ernzerhof (PBE) functional[51] and semi-empirical DFT-D3 van der Waals corrections[52] using the VASP code[53]. For each crystal, nonmagnetic (NM), ferromagnetic (FM), and three antiferromagnetic (AFM) configurations were considered.

The stability of these hypothetical crystals was examined by computing the formation energies and the phonon spectra using the phonopy package[54] and the VASP code. We first computed the formation energies ($E_{form}$) of each TMGT compound. Except for CuGT, which shows a positive $E_{form}$ value, the rest nine alloys are chemically stable with a $E_{form}$ ranging from –654.90 meV/atom (ScGT) to –23.30 meV/atom (FeGT). However, only five crystals, namely, ScGT, TiGT, CrGT, MnGT, and ZnGT, turn out to be dynamically stable according to the phonon dispersion calculations. Their crystalline (c-) structures and phonon dispersion curves are shown in Figure 2a and 2b, respectively. Among them, ScGT, TiGT and ZnGT are non-magnetic, while CrGT and MnGT are stable only when spin-polarization is considered. Large imaginary frequencies are found in the NM configuration for CrGT and MnGT. No matter whether NM or spin-polarized configurations are considered, the phonon curves of VGT, FeGT CoGT, NiGT, and CuGT show imaginary frequencies.

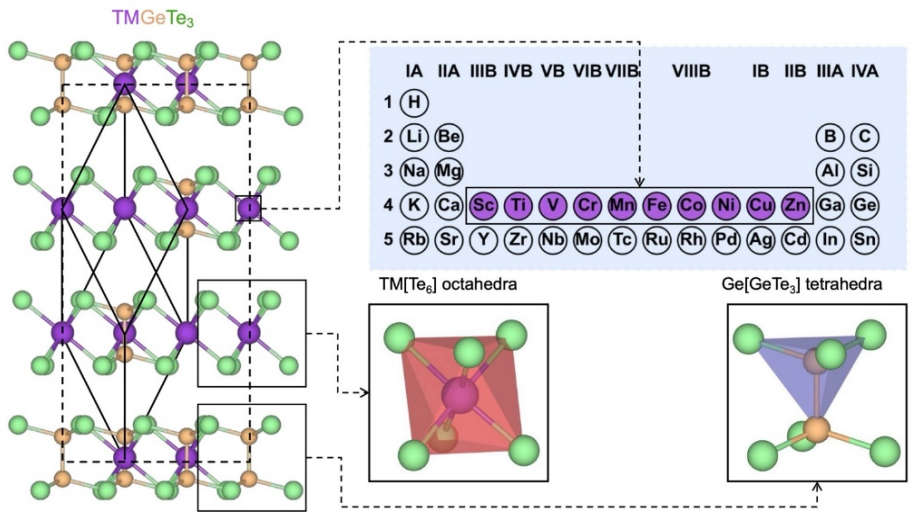

**Figure 1.** Atomic structure of crystalline TMGeTe$_3$ (TMGT) where TM denotes one of the ten 3$d$ transition-

metal elements shown in purple in the periodic table. Ge and Te atoms are rendered in brown and green respectively. The two insets on the lower right present the typical local structures, i.e. octahedra around TM atoms and tetrahedra around Ge atoms.

The DFT-relaxed lattice parameters and bond lengths are included in Figure 2a. The Ge-Ge and Ge-Te bonds are not much affected by the occupation of different TM elements, showing small variations of 2.43–2.46 Å and 2.61–2.64 Å. The TM-Te bonds show larger variations, ranging from 2.76 Å in c-CrGT up to 2.95 Å in c-ScGT. Their band structures are shown in Figure 2c. For c-ScGT, an indirect band gap of ~0.52 eV is observed near the Fermi level, indicating semiconducting behavior. With one more $d$ electron in c-TiGT, the Fermi level is shifted to higher energies and crosses several energy bands, resulting in metallic behavior. For c-CrGT, although more $d$ electrons are present, spin-polarization opens up the band gap again. The spin up and spin down channels are plotted in red and blue, respectively. In the NM case, there is no energy gap and a strong antibonding peak is found at the Fermi level, which seriously weakens the chemical stability[39]. c-MnGT is spin-polarized but metallic, since additional bands are partly occupied due to the presence of one more $d$ electron. Both spin up and spin down channels cross the Fermi level. Regarding c-ZnGT, the system turns to be non-magnetic because of the fully filled $d$ shell, and it shows metallic behavior.

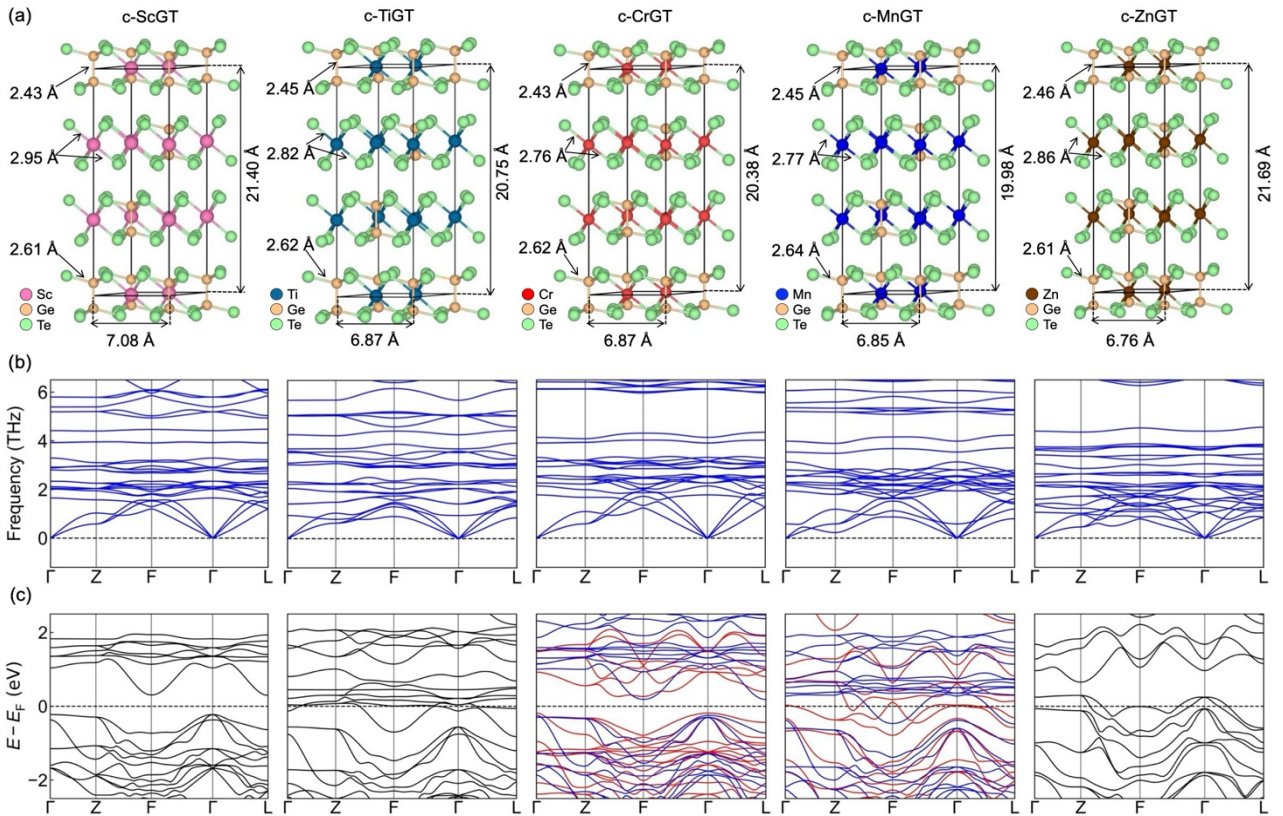

**Figure 2.** (a) The DFT-relaxed crystal structures of ScGT, TiGT, CrGT, MnGT, and ZnGT. The lattice parameters and bond lengths are noted. (b) Phonon spectrum for crystalline (c-) ScGT (non-magnetic, NM), TiGT (NM), CrGT (ferromagnetic, FM), MnGT (FM), and ZnGT (NM). (c) The band structures around the Fermi level of the five compounds. The bands are plotted in black for non-magnetic compounds, and in red (majority spin up states) and blue (minority spin down states) for magnetic compounds.

Next, we investigated the corresponding amorphous phase (a-) of these five alloys. DFT-based *ab initio* molecular dynamics (AIMD) simulations were conducted using the CP2K package[55] to generate amorphous models for these alloys following a melt-quenched scheme[56-60]. For each TMGT compound, 180 atoms (36 TM atoms, 36 Ge atoms, and 108 Te atoms) in a cubic box at the crystalline density were fully randomized at over 2000K for 30 ps and then temperature was decreased to ~1200K and the compound were held at this temperature for 30 ps to eliminate any remaining crystalline order. Then, the models were quenched with a rate of 12.5 K/ps to 300K, and were kept at 300 K for another 30 ps for data collection. During this quenching process, the box volume was gradually adjusted to keep its internal stress below 3 kbar. Figure 3a shows the obtained amorphous models of a-ScGT, a-TiGT, a-CrGT, a-MnGT, and a-ZnGT. The corresponding lattice edges are 17.73, 17.19, 17.10, 17.11 and 17.57 Å, respectively.

All amorphous models were further quenched from 300K to 0K within 24 ps and relaxed at 0K for electronic density of states (DOS) analysis (Figure 3b). The a-ScGT and a-ZnGT models show clear semiconducting features with a narrow energy gap of 0.39 eV and 0.48 eV, respectively, but a-TiGT shows no visible energy gap. For the spin-polarized a-CrGT and a-MnGT models, there are no energy gap for both spin channels. Hence, we predict a semiconductor-to-metal transition for ZnGT upon crystallization. As for ScGT, both phases are semiconducting, but the disordered nature of the amorphous phase might hinder the charge transport. Therefore, ScGT should display a sizable resistance contrast upon phase transition, similarly to GST. On the contrary, CrGT shows an inverse resistance contrast with the amorphous phase being the low-resistance state and the crystalline phase being the high-resistance one[33], due to the absence of an energy gap in the amorphous phase and the spin-polarization-induced opening of a band gap in the crystalline phase[39]. Regarding TiGT and MnGT, it is unclear whether a sizable contrast could be observed given the filled energy gap in both amorphous and crystalline phases. Nevertheless, the difference in the degree of disorder in the two phases should lead to some changes in carrier mobility.

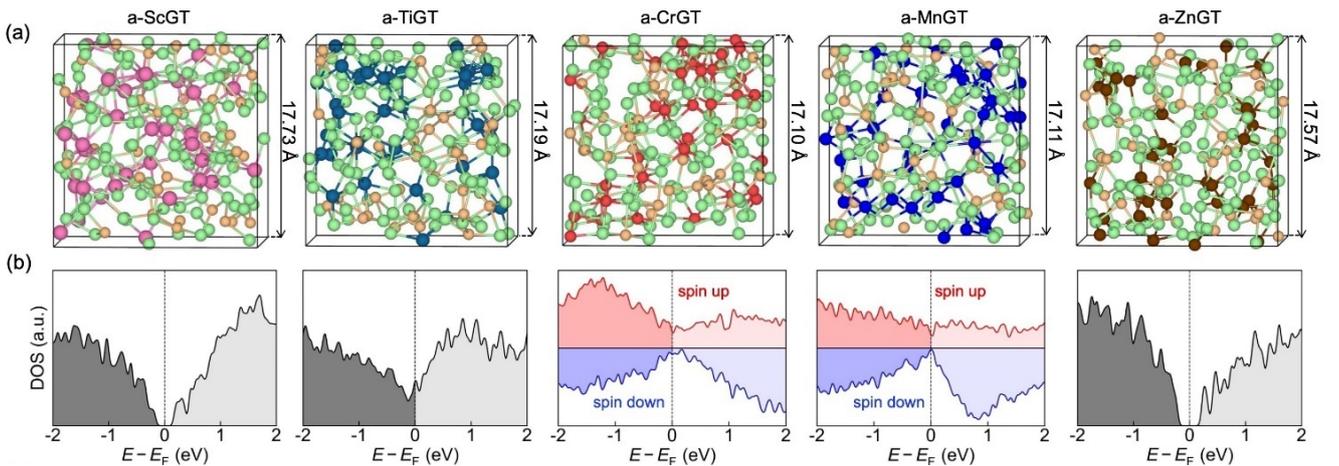

**Figure 3.** (a) The amorphous (a-) models of ScGT, TiGT, CrGT, MnGT, and ZnGT generated by the melt-quench simulation scheme. (b) The density of states (DOS) of the five amorphous models.

The partial radial distribution function (RDF) curves for the five amorphous models are displayed in in Figure 3c. The heteropolar TM-Te and Ge-Te bonds are the dominant bonding pairs in all five amorphous models with a major peak appearing at ~2.7–2.9 Å. Very limited fraction of short TM-TM bonds can be found below 3 Å for Sc-Sc and Zn-Zn. Instead, a stronger correlation between TM atoms is found at much larger interatomic distance from 3.5 to 4.5 Å. For the TM-Ge contacts, a small speak can be found below 3 Å for the five amorphous models, but this first peak is much weaker than the corresponding secondary peak at 3.5 Å and above. The homopolar Ge-Ge and Te-Te bonding pairs are in general similar in the five amorphous models. The angular distribution function (ADF) and coordination number (CN) distribution are shown in Figures 4a and 4b, respectively. For all the alloys except a-ZnGT, the TM atoms form primarily octahedral bonds with Te atoms showing a major ADF peak at ~90° and a secondary peak close to 170°. But for a-ZnGT, only one primary ADF peak is found at ~109° around Zn atoms and no secondary peak can be observed. For all the five amorphous models, the ADFs around Ge atoms and Te atoms show only one primary peak at ~90°. Regarding the CN distribution, Sc, Ti, Cr and Mn atoms have one to two more neighbors than Zn atoms. The overall CN distribution around Ge and Te atoms is similar for the five amorphous models with three to four neighbors. These observations indicate that the majority of Sc, Ti, Cr and Mn atoms form octahedral and 5-fold defective octahedral motifs but Zn atoms mainly form tetrahedral motifs.

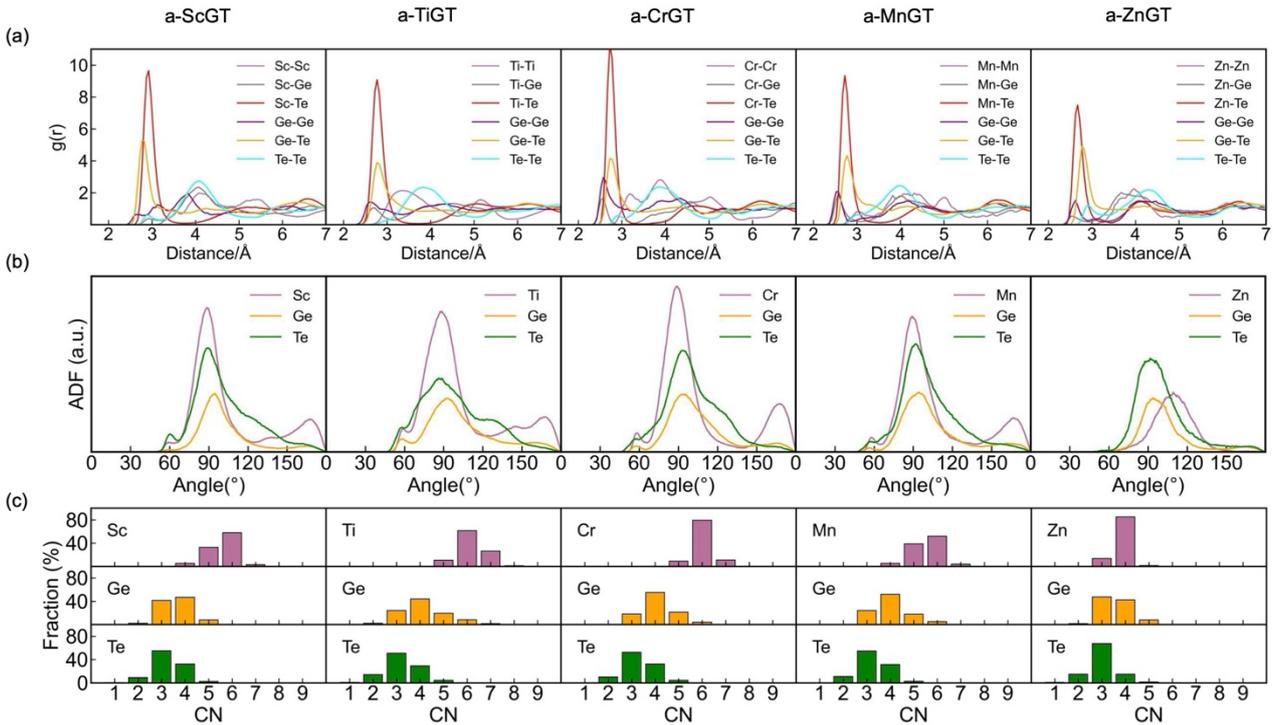

**Figure 4.** (a) The radial distribution function (RDF), (b) the angular distribution function (ADF), and (c) the distribution of coordination numbers (CN) calculated for the five amorphous models. The structural data are collected from the annealing trajectories at 300K for 30 ps and averaged over three independent melt-quenched amorphous models for each composition.

Given the spin-polarized nature of MnGT in both crystalline and amorphous phases, it is also interesting to investigate its potential for magnetic phase-change applications utilizing the magnetic contrast like in CrGT. In the following calculations, we primarily focus on ferromagnetic configurations, but also consider more complicated configurations with random distributions of spin-up and spin-down magnetic moments for comparison. Our aim here is to compare the magnetic properties of MnGT and CrGT compounds, since the latter is experimentally found to undergo a transition in both atomic structure and spin configuration to a glassy state. Therefore, it is interesting to investigate whether MnGT, which we predict to be stable, shows a similar transition to a spin-glass state.

Figure 5a presents the atomic magnetic moments distribution of Cr/Mn atoms in the respective amorphous models with the hypothetical FM configuration. In a-CrGT, the magnetic moments of Cr atoms are primarily found in the range of 2.88 to 3.60 $\mu_B$ with an average value of 3.14 $\mu_B$, which is comparable to the value of c-CrGT (3.11 $\mu_B$). Regarding MnGT, the magnetic moments of Mn atoms are more widely distributed, ranging from 2.90 to 4.06 $\mu_B$ with an average value of 3.53 $\mu_B$. This value is also close to the crystalline value (3.49 $\mu_B$). The fluctuations in local magnetic moments in the amorphous models can be attributed to the broad distribution of bond lengths and the presence of defective coordinations in the amorphous phase. Overall, the magnetic moment values of Mn atoms are higher than the those of Cr atoms. To model more complex magnetic configurations, we considered three random-magnetic (RM) models with half of the Mn atoms with positive magnetic moments (chosen in a random fashion), and the remaining ones with negative moments. We also considered three RM configurations for the other two amorphous MnGT models,. After relaxation, the nine RM models show only small net magnetizations (~0.005 $\mu_B$/Mn on average), but relatively low energy difference (2.17 meV/atom) with respect to their FM models. For a-CrGT, this energy difference was slightly larger, ~4 meV/atom[39]. This energy comparison indicates that a-MnGT might form a spin glass state more easily than a-CrGT.

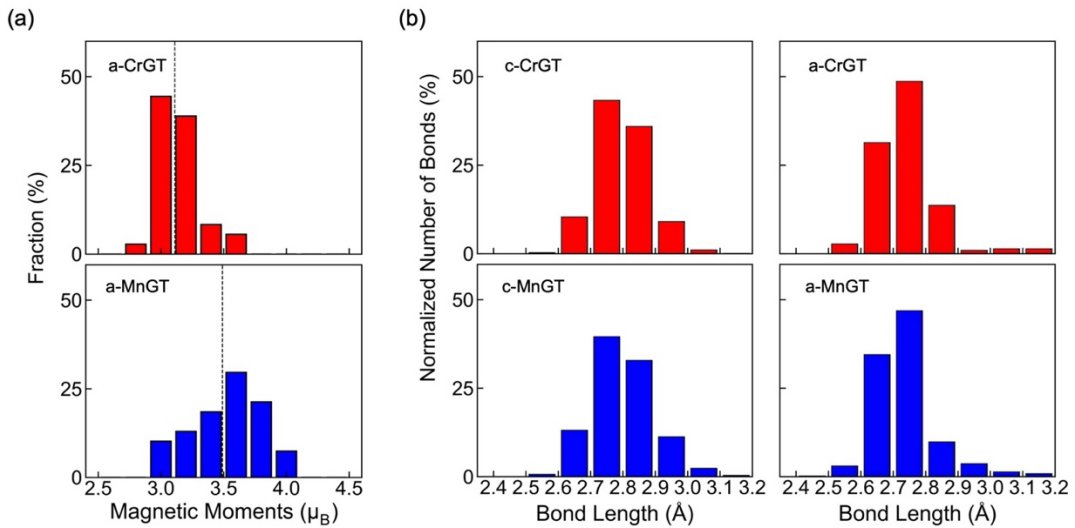

**Figure 5.** (a) Distribution of local magnetic moments in a-CrGT and a-MnGT models. The dashed lines mark the position of the local moment values in the respective crystalline models. (b) Distribution of the bond lengths between magnetic atoms and nearest neighbor atoms for Cr (red) and Mn (blue) atoms in amorphous and crystalline CrGT and MnGT models.

In our previous study, GST and GeTe alloys with Cr/Mn doping were predicted to show a magnetic contrast upon the transition from the crystalline to the amorphous state with a sizable difference in average magnetic moment by ~20%, which was explained by local structural changes around the magnetic impurities Cr and Mn[61,62]. Some follow-up magnetic experiments supported our predictions on ferromagnetic switching in Mn-doped GeTe[63,64]. However, for the Cr and Mn atoms in CrGT and MnGT, the concentration of magnetic atoms (20 at%) is much higher than in doped GST and GeTe (below 7 at%), and smaller differences between the short-range structure in crystalline and amorphous phases are found. To further compare the structural properties of the two phases, we firstly created a supercell for the relaxed crystalline models containing 360 atoms and annealed it at 300K for 30 ps for data collection. The bond length distribution of these supercells as well as those of the amorphous models are presented in Figure 5b. In both phases of CrGT and MnGT, the typical (40%~48%) lengths of the bonds involving Cr/Mn atoms are consistently found in the range of 2.7~2.8 Å, in contrast with the bonding scenario in Cr/Mn doped GST, where a major shift in bond length distribution was found upon amorphization. The average bond length around Cr and Mn atoms was reduced by 2.04% and 2.36 % upon amorphization, and the stronger *p-d* hybridization gave rise to an obvious reduction in local magnetic moments for amorphous Cr- and Mn-doped GST [61].

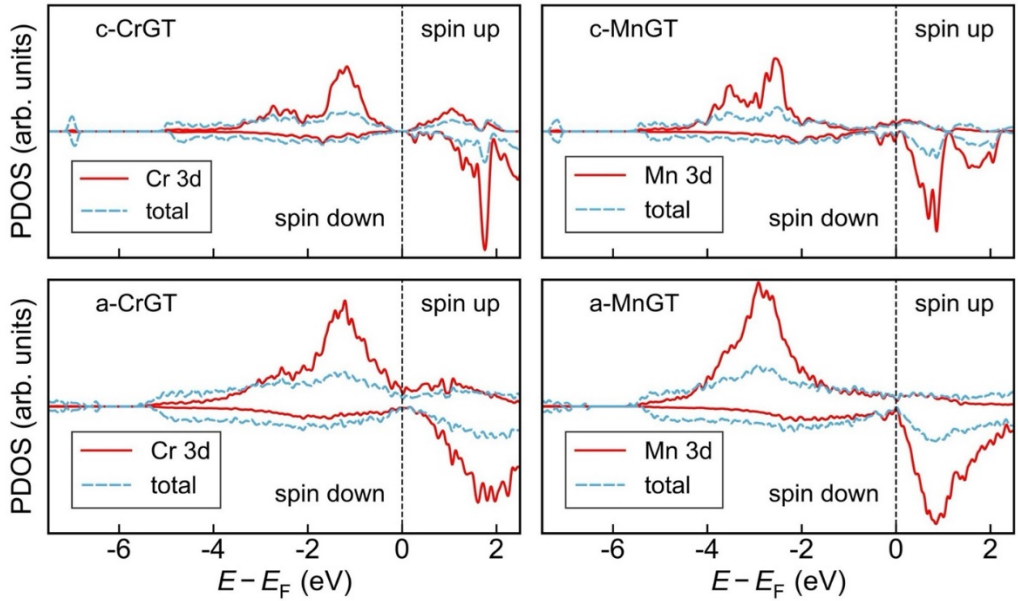

**Figure 6.** Electronic structure analysis of the c-phase and a-phase of CrGT (left) and MnGT (right). The dashed and solid lines indicate the total DOS and the DOS projected onto the 3*d* states of the magnetic atoms (PDOS) respectively. The profiles of the 3*d* states are enlarged for better visibility.

The projected DOS on the 3*d* orbitals of the Cr/Mn atoms for CrGT and MnGT in both crystalline and amorphous phases are shown in Figure 6. In c-CrGT, the majority spin states of the Cr 3*d* orbitals lie mainly below $E_F$, while the minority spin states are almost unoccupied and located at much higher energies. Given that c-CrGT is a narrow band semiconductor, this suggests that the FM order in the system could be stabilized by carrier-mediated *p-d* exchange interactions, in addition to the Cr-Te-Cr superexchange mechanism[50]. Regarding c-MnGT, the profiles of majority and minority spin states are fairly similar to c-CrGT, but the positions are shifted downwards. Nevertheless, the minority spin states still lie mainly above $E_F$, whereas the majority spin states are mostly occupied, resulting in larger magnetic moments compared to c-CrGT. This trend suggests that c-MnGT shares a similar exchange mechanism for the FM order. Upon amorphization, both compounds show broader distributions, while the position of the major peaks remains unshifted. Besides the 20% change in average magnetic moment in Cr/Mn doped GST upon amorphization, CrGT and MnGT could also undergo a major magnetic phase transition, i.e. from FM to spin glass. This drastic change in magnetic response induced by the structural transition could be exploited for spintronic applications, where fast magnetic switching is desired. Further magnetic and transport experiments are anticipated to evaluate the predicted properties of crystalline and amorphous MnGT alloys.

**Conclusion**

In summary, a systematic investigation of 3$d$ transition metal substituted CrGT-like phase change materials was performed. In addition to CrGT, four TMGT alloys, i.e., ScGT, TiGT, ZnGT and MnGT, were predicted to be chemically and dynamically stable in the CrGT-like layered structure. Among them, ScGT, TiGT and ZnGT are nonmagnetic, while MnGT is ferromagnetic in the crystalline form. ScGT shows narrow-gap semiconducting properties in both phases, indicating that it might exhibit conventional resistance contrast. For ZnGT, the typical local structures around Zn atoms change from octahedral configurations in the crystalline phase to tetrahedral motifs in the amorphous phase. Upon amorphization, ZnGT is predicted to undergo a metal-to-semiconductor transition, showing a much larger resistance window. It is difficult to predict whether there is a resistance contrast for TiGT, because there is no energy gap in either crystalline or amorphous phases. Nevertheless, the degree of disorder should lead to some difference in carrier mobility. The magnetic nature of MnGT can be preserved even in the amorphous phase. In general, the magnetic moments in MnGT are slightly larger than in CrGT in both phases, and it is predicted that the tendency towards the formation of a spin-glass state in the amorphous phase at low temperatures could be even stronger than in CrGT. Therefore, MnGT is a promising candidate for magnetic phase-change applications, similarly to CrGT. Given the complexity of seed formation, these four TMGT alloys should all have a high crystallization temperature. We expect our work to inspire further efforts in identifying more materials candidates for embedded phase-change memory and fast magnetic switching applications.

**Methods**

The VASP code[53] was employed to perform DFT calculations including structural relaxation and self-consistent calculations for both crystalline and amorphous models. The projector augmented wave (PAW) method[65] and the PBE functional[51] were used with an energy cutoff of 450 eV. The k-point meshes for crystalline and amorphous models were 5×5×5 and gamma-only, respectively. The phonon spectrums were computed using the VASP software and phonopy package[54]. To generate amorphous models, AIMD simulations using the CP2K package[55] were performed. For these simulations, Goedecker pseudopotentials[66], and the PBE functional were employed. The Kohn–Sham orbitals were expanded in a Gaussian-type basis set with double-/triple-ζ polarization quality, whereas the charge density was expanded in plane waves, with a cutoff of 300 Ry. The AIMD calculations were performed using the second-generation Car-Parrinello scheme[67] in the canonical (NVT) ensemble with a time step of 2 fs. The spin polarization was achieved using $\alpha$ and $\beta$ orbitals without any spin restriction during the AIMD simulations. The semi-empirical DFT-D3 van der Waals corrections[52] were included for all the calculations. The Mulliken charge analyses were performed using LOBSTER[68].

**Data Availability Statement**
Data supporting this work will be available at https://caid.xjtu.edu.cn/info/xxxxx upon journal publication.

**Competing interests**
The authors declare no competing interests.

**Acknowledgments**
The work is supported by the National Key Research and Development Program of China (2023YFB4404500). J.-J.W. thanks the support of National Natural Science Foundation of China (62204201). W.Z. thanks the support of National Natural Science Foundation of China (62374131). We acknowledge the HPC platform of Xi'an Jiaotong University and Computing Center in Xi'an for providing computational resources. The authors acknowledge XJTU for hosting their work at CAID. The International Joint Laboratory for Micro/Nano Manufacturing and Measurement Technologies of XJTU is acknowledged. R. M. gratefully acknowledges funding from the PRIN 2020 project "Neuromorphic devices based on chalcogenide heterostructures" funded by the Italian Ministry for University and Research (MUR).